\documentclass[preprint,aip,jcp,superscriptaddress]{revtex4-1}

\usepackage{graphicx,color}
\usepackage{amsmath,amsfonts,amssymb}
\usepackage{graphicx}
\usepackage[caption=false]{subfig}
\usepackage{dcolumn}
\usepackage{placeins} % ADDED FOR FLOAT BARRIERS IN DRAFT

\newcommand{\fig}{Fig.}

\newcommand{\figref}[1]{\fig~\ref{#1}}

\newcommand{\tabref}[1]{table~\ref{#1}}

\renewcommand{\eqref}[1]{equation~(\ref{#1})}

\newcommand{\rxnref}[1]{reaction~(\ref{#1})}

\newcounter{defcounter}
\newenvironment{reaction}{%
\addtocounter{equation}{-1}
\refstepcounter{defcounter}

\begin{equation}}
{\end{equation}}

\newcommand{\new}[1]{#1} % i.e. do nothing.

\begin{document}

\title{Potential energy surface interpolation with neural networks for instanton rate calculations}

\author{April M. Cooper}
\affiliation{Institute for Theoretical Chemistry, University of Stuttgart, Pfaffenwaldring 55, 70569 Stuttgart,Germany}

\author{Philipp P. Hallmen}
\affiliation{Institute for Theoretical Chemistry, University of Stuttgart, Pfaffenwaldring 55, 70569 Stuttgart,Germany}

\author{Johannes K\"{a}stner}
\affiliation{Institute for Theoretical Chemistry, University of Stuttgart, Pfaffenwaldring 55, 70569 Stuttgart,Germany}

%\date{\today}

\begin{abstract}
Artificial neural networks are used to fit a potential energy surface. We demonstrate the benefits of using not only energies, but also their first and second derivatives as training data for the neural network. This ensures smooth and accurate Hessian surfaces, which are required for rate constant calculations using instanton theory. Our aim was a local, accurate fit rather than a global PES, because instanton theory requires information on the potential only in the close vicinity of the main tunneling path. Elongations along vibrational normal modes at the transition state are used as coordinates for the neural network. The method is applied to the hydrogen abstraction reaction from methanol, calculated on a coupled-cluster level of theory. The reaction is essential in astrochemistry to explain the deuteration of methanol in the interstellar medium. 
%Abstract of roughly 250 words
\end{abstract}

%\keywords{}

\maketitle

%%%%%%%%%%%%%%%%%%%%%%%%%%%%%%%%%%%%%%%%%%%%%%%%%%%%%%%%%%%%%%%%%%%%%%%%%%%%%%%%%%%%%%%%%%%%%%%%%%%%%%%%%%%%
\section{Introduction}

Reliable information on the potential energy surface (PES) is crucial to many different applications like molecular dynamics (MD), Monte Carlo (MC) or quantum dynamics simulations. Further, a correct description of the PES is of great importance for the calculation of  reaction rate constants,\cite{mei16} e.g. with the instanton method.
Therefore, accurate methods to obtain information on the PES are of great interest.
In principle it is possible to calculate the information on the PES
needed by performing ab initio energy calculations on-the-fly during
the simulation process, \new{i.e. whenever information on the PES is
  needed it is calculated by electronic structure theory.} However,
this procedure increases the time needed for the simulation
significantly as the computational demand for these electronic
structure calculations, \new{from which the energy and its derivatives are obtained,} is usually very high. Therefore, this approach is often infeasible. 
 
In order to avoid this additional computational cost during the simulation,
one can precompute the PES by an interpolation of ab-initio data and then use
the information on the PES during the desired simulation application. Several
approaches have been used to precompute PESs like spline interpolation,
\cite{cha83,rei84,bow86,pre07} modified Shepard interpolation
\cite{lan86,far87} or interpolating moving least squares.
\cite{ish99,mai03,guo04,daw07}  Another very promising approach is the
interpolation of ab initio data with the help of artificial neural networks
(NNs) to determine potential energy surfaces. This approach has become popular
during the last two
decades.\cite{bla93,gas93,bla95,taf96,bro96,no97,gas98,pru98,pru98a,mun98,hob99,cho02,roc03,bit04,lor04,wit05,man06,man06b,man06a,agr06,dou06,mal07,le09,puk09,han10,beh10,che13,
  yua15}

The interpolation of ab initio data with a NN allows to precompute the PES in
a very accurate manner, with the result that during the actual simulation
practically no computational effort is needed to request any information on
the PES.  Neural networks are well-suited for this task because they do not
restrict the interpolation of the PES to a specific functional form. It was
formally proven that artificial neural networks are universal
approximators.\cite{man06,lud07,le10} 

At present it is common practice to use
only the energy during the training process of the NN. However, it was
suggested in several publications \cite{fer05,wit05,puk09,beh11,art11} to add
information on the gradient of the energy with respect to the input
coordinates. To our knowledge the use of Hessian information in the NN fit of
a PES has \new{only been reported rarely yet.}
\new{In the field of minimum energy path (MEP) searches,\cite{pet16,kho16,pet17,koi17} it was shown that including Hessian information could lower the number of iterations needed to obtain a reliable MEP employing Gaussian process regression.\cite{koi17}}
This approach \new{of including the gradient and Hessian information in the NN fit} improves the quality of forces
that are obtained from the interpolated surface because it ensures that not
only the energy itself but also its gradient and Hessian are accurately
fitted.  For the calculation of reaction rate constants with instanton theory
it is important to ensure that the energy, as well as the elements of the
gradient and especially the Hessian are smooth functions of the input
coordinates. Unfortunately, standard methods, like \new{standard} Shepard interpolation,
\cite{far87} often lead to spikes in the hyper surfaces describing the Hessian
matrix elements and are, thus, not well-suited for the calculation of reaction
rate constants with instanton theory. \new{However, more recent variants of
  Shepard interpolation reduce these issues.\cite{bet99,yan01,col02,fra14}}
Since NNs are universal approximators
for smooth functions, the Hessian matrix elements can be fitted such that the
resulting hyper surfaces describing the change of these with respect to the
input coordinates are smooth. Therefore, we suggest an approximation of the
PES by fitting a NN to information on the potential energy, gradients and
Hessians in an explicit manner.  

Incorporating gradient and Hessian information in the NN training increases
the computational effort for the generation of each control point of the
training and test sets. Nevertheless, this computational effort is feasible
for the calculations of reaction rate constants where a fit to only a local
part of the PES is necessary, in contrast to the global fits required for MD
or MC simulations. For the calculation of rate constants with instanton theory
it is sufficient to fit a comparatively small sector of the PES in the
proximity of the first order saddle point that corresponds to the transition
state of the reaction of interest. Further, it should be ensured that the
proximity of the reactant state minimum is well described. Since the PES has
to be described only locally, it is sufficient to choose comparatively few
molecular configurations for the construction of the training and test set.
This implies that a manageable number of ab initio calculations need to be
done to allow for a good fit of the PES, which keeps the overall computational
effort within reasonable bounds.

The method is applied to calculate rate constants for hydrogen abstraction from methanol by an incoming hydrogen atom, 
\begin{reaction}
  \text{CH}_3\text{OH}+\text{H} \rightarrow
  \text{CH}_2\text{OH}+\text{H}_2.
  \label{rkn:1}
\end{reaction}
This reaction is crucial in astrochemistry to explain the high degree of deuteration of methanol in dense clouds in the interstellar medium.\cite{gou11}

This paper is organized as follows. We give a brief overview of instanton
theory to clarify why a small sector
of the potential energy surface is sufficient to calculate rate constants, but
why accurate Hessian information in that sector is required. Next, we
describe details of the NN setup and how we use energy, gradient, and Hessian
information to train the network. Then we explain how we sample the
configurational space to define the control points for the training
data. Next, the averaging process of independently trained NNs and the
resulting error measure are described. In Results and Discussion we apply the
theory to \rxnref{rkn:1} and calculate rate constants with different settings.

%%%%%%%%%%%%%%%%%%%%%%%%%%%%%%%%%%%%%%%%%%%%%%%%%%%%%%%%%%%%%%%%%%%%%%%%
\section{Methods}
%%%%%%%%%%%%%%%%%%%%%%%%%%%%%%%%%%%%%%%%%%%%%%%%%%%%%%%%%%%%%%%%%%%%%%%%

\subsection{Instanton calculations}

Semiclassical instanton theory\cite{lan67, lan69, mil75, cal77, gil77, aff81,
  col88, han90, ben94, mes95, ric09, kry11, alt11, rom11, rom11a, kry14,
  zha14, ric16} provides a way to calculate rate constants including atom
tunneling based solely on geometry optimizations, i.e. without dynamical
sampling. Thus, the computational effort is kept at bay. The rate constant is
obtained as the imaginary part of the (logarithm of the) partition
function. The latter is calculated in Feynman path integral
formulation\cite{fey48} using the steepest descent approximation of the phase
space integral.\cite{kae14} In that way, the most likely tunneling path at a
given temperature, the instanton, has to be located. The partition function is
then calculated by taking fluctuations around the instanton path into account
to quadratic order. The traditional way to find an instanton was to locate a
periodic orbit on the inverted potential energy surface.\cite{ceo12} It is
much more efficient, however, to discretize the Feynman path and search a
first-order saddle point in the space of discretized Feynman paths.\new{\cite{and09,rom11}} A
quadratically converging search algorithm\cite{rom11a} allows to locate
instantons efficiently in high-dimensional systems.

To calculate the rate constant $k_\text{inst}$, fluctuations around the
instanton path are taken into account, leading to \cite{rom11a}
\begin{equation}
  k_\text{inst}=\sqrt{\frac{S_0}{2\pi\hbar}} \sqrt{\frac{P}{\beta\hbar}} 
  \frac{\prod_{l=N_0+1}^{NP}\sqrt{\lambda_l^\text{RS}}}
       {\prod_{l=N_0+2}^{NP}\sqrt{|\lambda_l^\text{inst}|}}
  \exp(-S_\text{E}/\hbar)
  \label{eq:kinst}
\end{equation}
Here, $N$ is the number of degrees of freedom, \new{$N_0$ is the number of translational and rotational degrees of freedom,} $P$ is the number of
discretization points of the Feynman path (images), $\beta$ is the inverse
temperature, $\beta=1/k_\text{B}T$, $\hbar$ is Planck's constant, $S_\text{E}$
is the Euclidean action of the instanton path and $S_0$ its shortened
action.\cite{rom11} The values $\lambda_l^\text{inst}$ and
$\lambda_l^\text{RS}$ are the eigenvalues of the second derivative matrix of
the Euclidean action of the instanton and the reactant state, respectively,
with respect to all coordinates of all images:
\begin{equation}
  \frac{\partial^2S_\text{E}}{\partial y_k^a\partial y_l^b}=
  \frac{P}{\beta\hbar}\delta_{a,b} (2\delta_{k,l}-\delta_{k-1,l}-\delta_{k,l-1}) +
  \frac{\beta\hbar}{P}\delta_{k,l} \frac{\partial^2E}{\partial y_k^a\partial y_k^b}
  \label{eq:hse}
\end{equation}
where $y_k^a$ is the mass-weighted coordinate component $a$ of image
$k$. Consequently, $\frac{\partial^2E}{\partial y_k^a\partial
  y_k^b}=\nabla_k\nabla_k E$ is the
second derivative (Hessian) of the potential energy of image $k$. Thus,
Hessians of all images along the instanton path are required to calculate rate
constants. The evaluation of those by on-the-fly calculations is typically the
most time-consuming step during an instanton calculation. Smooth and accurate
Hessians are a pre-requisite for reliable rate calculations on fitted
potential energy surfaces. This is why we use them to train our NN-PES.

Translation and rotation can be taken into account separately in
$k_\text{inst}$ by assuming them to be decoupled from the vibrations treated
in \eqref{eq:kinst}.

In this paper the Feynman paths were discretized to $P=200$ images and the instantons were optimized such that the
gradient of $S_\text{E}$ with respect to the mass-weighted coordinates was
smaller than $5.0 \cdot 10^{-11}$ atomic units. Such a small threshold is
generally only achievable for PESs and derivatives with negligible numerical noise.

\subsection{Neural Network Setup}

In order to obtain a correct description of the local PES needed for the determination of the reaction rate constant, a NN is trained to predict the potential energy $E$ that corresponds to a given configuration defined by the input coordinates $x_i, \, i =1,\dots,I$. 
For this purpose we use a feed-forward neural network with 2 hidden layers and a single output node for the potential energy. Thereby the nodes in layer $l$ are connected to every node in layer $l+1$. The output of the two hidden layers $y_j^1,y_k^2$ and the potential energy $E=y$ were calculated as follows:
\begin{align}
	y_j^1 &= f^1\left(b_j^1+\sum_{i=1}^I\left(w_{j,i}^1 \cdot x_i\right)\right), \quad j=1,\dots , J \\
	y_k^2 &= f^2\left(b_k^2+\sum_{j=1}^J\left(w_{k,j}^2 \cdot y_j^1\right)\right), \quad  k=1,\dots , K \\
	E&=y_{\text{\tiny{NN}}}=f^3\left(b_1^3+\sum_{k=1}^K\left(w_{1,k}^3 \cdot y_k^2\right)\right)
\end{align}
where $I$ is the number of nodes in the input layer (number of input coordinates) and $J,K$ are the number of nodes in the first ($J$) and second ($K$) hidden layer. Further is $w_{b,a}^l$ the weight connecting node $a$ in layer $l$ with node $b$ in layer $l+1$. Whereas $b_a^l$ is the bias acting on node $a$ in layer $l$. The transfer functions are denoted $f^l,\, l\in \{1,2,3\}$.   

The transfer functions $f^1$ and $f^2$ are chosen as ${f^1(\cdot)=f^2(\cdot)=\tanh(\cdot);}$
$f^3$ is chosen to be $f^3(x)=x$ in order to allow for the prediction of arbitrary potential energy values.

In order to be able to fit not only the energy, but also the gradient and
Hessian with respect to the input coordinates by a NN, these two quantities
are calculated from the potential energy predicted by the NN. Thereby all
emerging derivatives are calculated analytically during a backward pass
through the NN by applying the chain rule. This is done to ensure that the
gradient and Hessian are the analytic derivatives for a given NNPES. In
principle it would also be possible to include the gradient or Hessian in the
output of the NN. If these quantities would also be fitted directly by the NN,
the resulting gradient and Hessian would contain small fitting errors. This,
however, implies, that they are not given by the exact analytical derivatives
of the potential energy and thus wouldn't be completely consistent with the energy
hypersurface. Further, it is better to calculate the gradient and Hessian
analytically from the energy predicted by a NN in order to keep the
computational effort of the NN training at bay. Including those quantities in
the NN output would require a significantly more complex structure of the NN
in order to allow for sufficient flexibility for the fit. Unfortunately the
computational demand of the NN training increases strongly with the number of
parameters that need to be optimized during the training, i.e. the number of
weights and biases. Thus, we calculate $\mathbf{g}_{\text{\tiny{NN}}}=\nabla
E$ and $\mathbf{H}_{\text{\tiny{NN}}}=\nabla\nabla E$ analytically. The
corresponding equations are provided in the Supporting Material.
%This, however, implies that the training process becomes
%very time consuming if the energy, as well as the gradient and Hessian shall
%be predicted directly by the NN with the same accuracy as in the approach
%where only the energy is fitted by the NN and the derivatives are calculated
%analytically.

In order to measure the quality of a NN fit a cost function has to be introduced.
Commonly the mean square error of the potential energy predicted by the NN is  used as a cost function for NN fits of PESs. As not only energies, but also gradients and Hessians are to be fitted by the presented procedure, this idea of taking the mean square error of the quantities that the NN is fitted to is extended to the elements of the gradient and Hessian. Thus, the cost function $R$ used in the training process is:

%\begin{align}\label{eq:costfct}
\begin{multline}
	R=\frac{1}{N_\mathrm{E}+N_\mathrm{G}+N_\mathrm{H}} \left[A_\mathrm{E}\sum_{e=1}^{N_E} (y_{\text{\tiny{NN}},e}-E_{\text{ref},e})^2 \right.  \\
	\left. +A_\mathrm{g}\sum_{g=1}^{N_G}\left| \mathbf{g}_{\text{\tiny{NN}},g}-\mathbf{g}_{\text{ref},g}\right|^2 					
	+A_\mathrm{H}\sum_{h=1}^{N_H}\left|\mathbf{H}_{\text{\tiny{NN}},h}-\mathbf{H}_{\text{ref},h}\right|^2 \vphantom{\int_1^2} 		
	\right]. \label{eq:costfct}
\end{multline}
%\end{align}

Thereby $\left| \mathbf{g}_{\text{\tiny{NN}},g}-\mathbf{g}_{\text{ref},g}\right|^2$ and $\left| \mathbf{H}_{\text{\tiny{NN}},h}-\mathbf{H}_{\text{ref},h} \right|^2$ are to be understood element wise, e.g.
\begin{equation} 
	\left| \mathbf{H}_{\text{\tiny{NN}},h}-\mathbf{H}_{\text{ref},h} \right|^2 \widehat{=} \sum_{m,n=1}^I \left| 		(\mathbf{H}_{\text{\tiny{NN}},h})_{mn}-(\mathbf{H}_{\text{ref},h})_{mn} \right|^2
\end{equation}

The weighting parameters $A_\text{E},\,A_\text{g}$ and $A_\text{H}$ quantify the relative influence of the errors in the energy, gradients and Hessians on the residual $R$. This introduces a measure of relative importance by penalizing errors in certain quantities more than in others. 

During the NN training derivatives of the cost function with respect to the
weights and biases need to be calculated, i.e. expressions for $\partial
E/\partial w_{b,a}^l, \, \partial E/\partial b_{a}^l $ but also for $\partial
\mathbf{g}/\partial w_{b,a}^l, \, \partial \mathbf{g}/\partial b_{a}^l$ and
$\partial \mathbf{H}/\partial w_{b,a}^l, \, \partial \mathbf{H}/\partial
b_{a}^l$ are required, which are tedious but straight forward to derive. These
derivatives are calculated analytically in our code to ensure a proper
optimization of the weights and biases. The specific expressions are provided
in the Supporting Material.
The cost function $R$ is minimized using an L-BFGS algorithm.\cite{liu89}

Further, it can be utilized that the NN output, i.e. the potential energy, as well as the corresponding gradient and Hessian are linear in the bias and weights of the third layer, i.e. $b_1^3$ and $w_{1,k}^3$. Therefore, it is possible to first do a linear optimization of $w_{1,k}^3,b_1^3$ and subsequently, given these weights and biases of the third layer, optimize the remaining weights and biases with a non-linear optimization method, like the L-BFGS algorithm. This was done in the calculations presented in this paper in order to accelerate the training process. 

The quality of the NN output depends strongly on the coordinate definition of
the input values. Therefore, Cartesian coordinates are not well-suited as input
for the NN as the NN output would change if the whole system would be
translated or rotated as the system's coordinates change. Thus, it is
important to choose a coordinate description that is invariant with respect to
rotations and translations. This is the reason why PESs are often fitted by
NNs using internal coordinates. Since the neural network allows for a
description of the input that contains redundant information it is even
possible to use the set of all interatomic distances to describe the input
structures.\cite{dou06} It is also possible to include further symmetry of the
system like the interchangeability of atoms of the same species or geometrical
symmetry of the structures. \cite{beh11a,smi17} We use mass-weighted
elongations along normal modes to describe the geometry of the NN input
structures as these are invariant with respect to translation and rotation of
the whole structure and describe the geometry uniquely. To transform from
Cartesian coordinates to a normal mode description, first, the structure is
superimposed with the transition state structure, which is used as the
reference structure, to eliminate any relative rotations and translations of
these two structures. After that the elongations of the structure with respect
to the transition state structure are described by an expansion in normal
modes. 

To test our approach for the calculation of the reaction rate constant we fitted a PES in the proximity of the reaction path of the reaction CH$_3$OH + H $\rightarrow$ CH$_2$OH + H$_2$ by a NN. Using the resulting PES for the energy as well as the gradient and Hessian matrix elements, we calculate the reaction rate constants for this reaction with the instanton method.  

Thereby NNs with the structure 15-50-50-1 are used, i.e. there are 15 input
nodes, 50 nodes in each hidden layer and one output node. The weights and
biases are initialized with uniformly distributed random numbers in the
interval $[-0.5,0.5]$, $b_1^3$ is initialized as the average of the energies
that correspond to the training set structures. The NN training is done over
5000 epochs after which we found $R$ to be sufficiently converged. The weight factors in the cost function $R$ are, after a preliminary parameter study, chosen to be $A_E=1.0, \, A_g=0.1 ,\, A_H=5.0$. All parameters are given in atomic units.

\subsection{Generation of the training and test data}

\begin{figure*}[t]
  \centering
  \includegraphics[width=17cm]{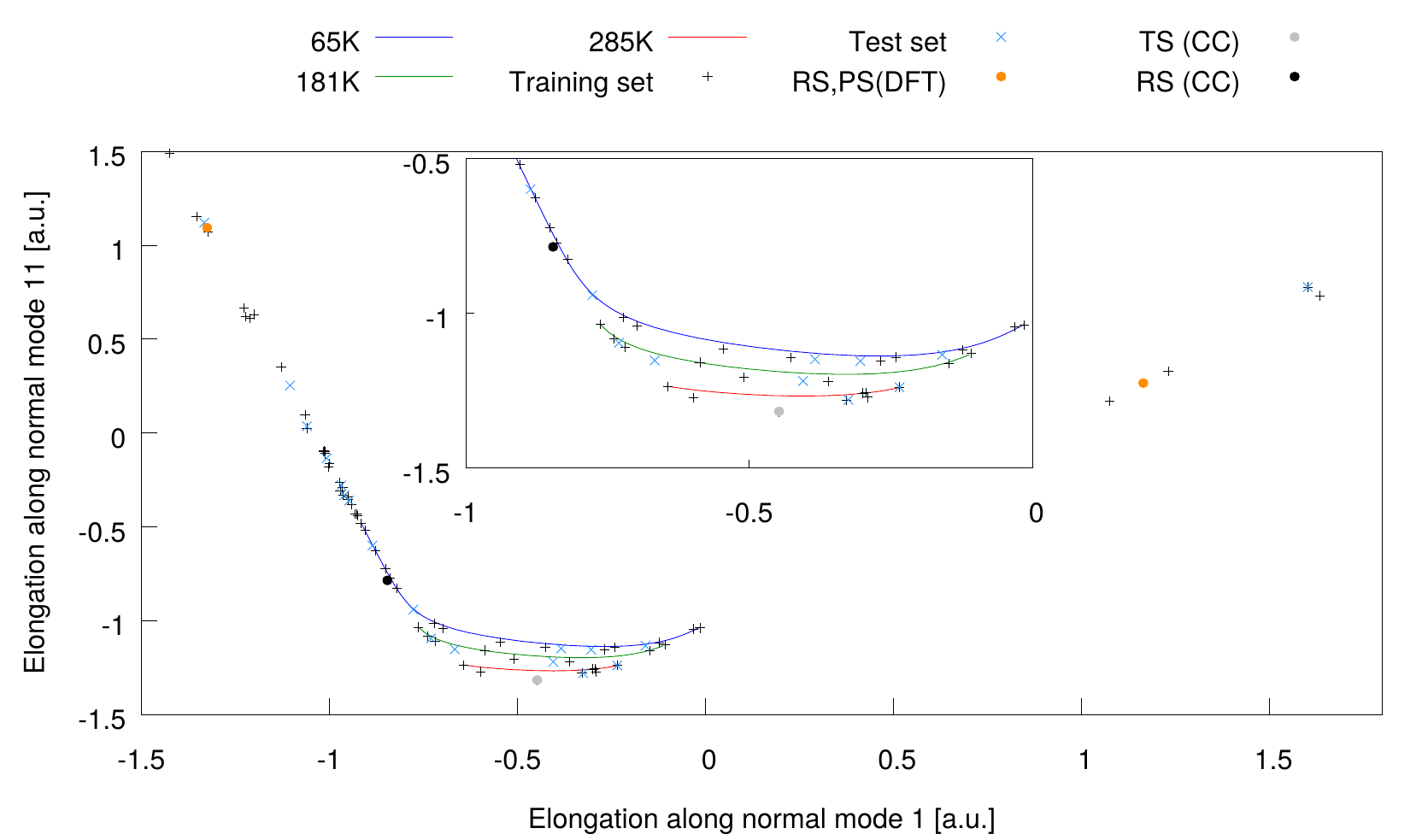}
  \caption{Instanton paths (lines) and structures used in the NNPES fitting process. Coordinates are given by elongations along normal modes 1 and 11. Stationary points on the different levels of theory are indicated by filled circles.}
  \label{fig:instantons}      
\end{figure*}

\begin{figure}
  \includegraphics[width=5cm]{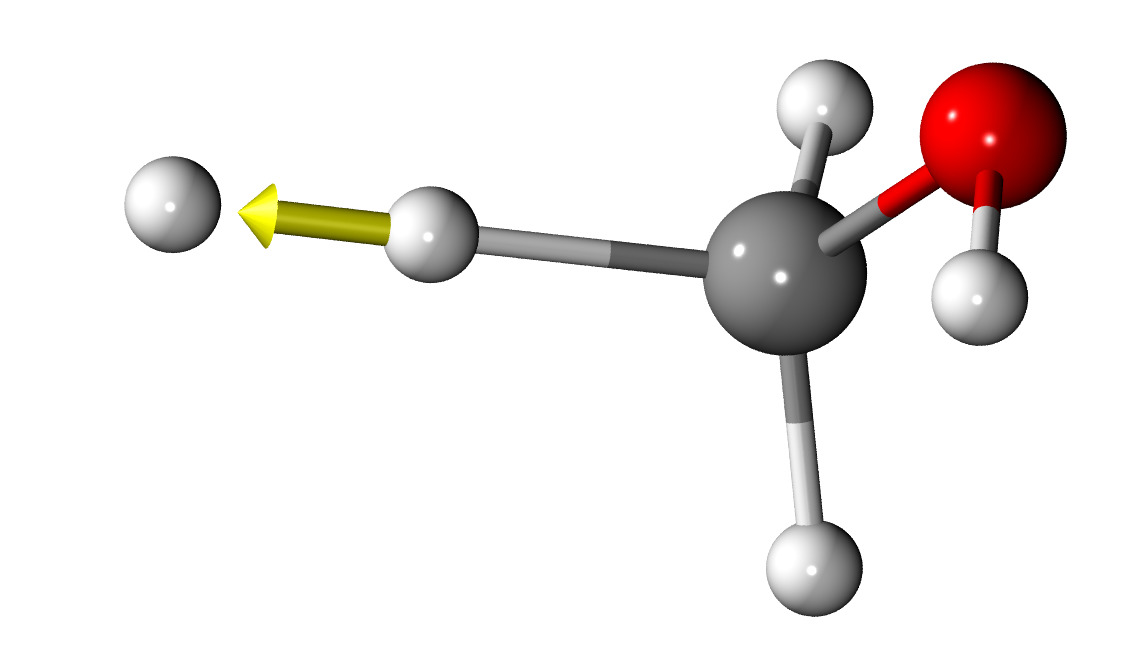}\\
  \includegraphics[width=5cm]{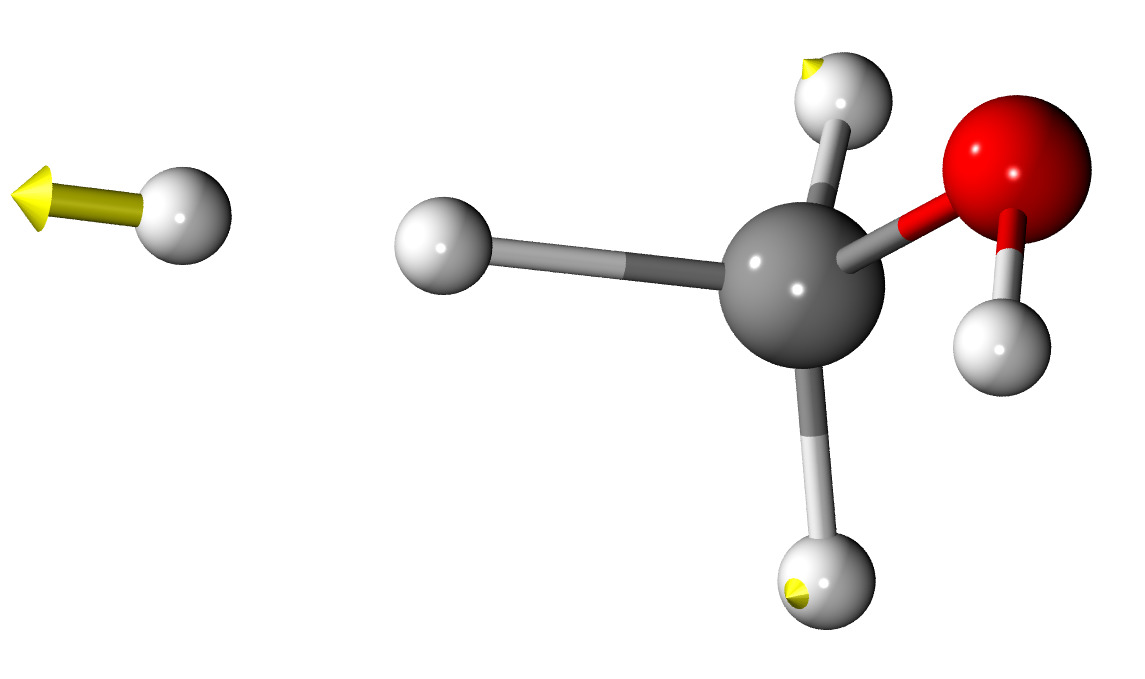}
  \caption{Normal modes 1 and 11, on which the geometries are projected to result in \figref{fig:instantons}.}
  \label{fig:modes}
\end{figure}
The purpose of the NN training is to define a local NNPES that can be used for
the calculation of reaction rate constants of one specific reaction.
Therefore, the molecular structures for which the energies, gradients and
Hessians were included in the training and test set were chosen along
instantons at different temperatures. The chosen target level of theory is
unrestricted explicitly correlated coupled-cluster theory including single and
double excitations and triple-excitations approximated perturbatively,
UCCSD(T)-F12/VTZ-F12, on a restricted Hartree--Fock (RHF)
basis.\cite{adl07,pet08,kni09} Gradients and Hessians were obtained by finite
differences of energies. However, calculations of instantons on
UCCSD(T)-F12/VTZ-F12 level, that require on-the-fly energy calculations, are
computationally barely feasible for the reaction at hand. Thus, geometries
along two instanton paths that were calculated with density functional theory
on BB1K/6-311+G** level of theory\cite{zha04a} at 285K and 200K
were used as a starting point. This level of theory was chosen because the
required computational demand is manageable and was previously\cite{gou11} found to result in a classical transition state structure 
close to the one obtained from UCCSD(T) calculations extrapolated to the
complete basis set. Since the region of the PES that is close to the
transition state is predominantly influencing the reaction rate, a correct
description of the classical transition state and its direct surroundings on
the PES is crucial.  The DFT calculations were performed in
ChemShell\cite{she03,met14} using NWchem.\cite{val10}
 
The training  and test set were created iteratively. 
Initially from each of the two DFT instanton paths 20 geometries were chosen. Thereby the structures obtained from the instanton path at 285K were used as an initial training set, the others as an initial test set. Further, the geometries of the pre-reactive van-der-Waals minimum and the classical transition state were added to the training set.
For every chosen geometry an ab initio calculation of the energy, the gradient and the Hessian matrix with respect to the spatial coordinates was performed on UCCSD(T)-F12/VTZ-F12 level of theory using Molpro 2012.1\cite{wer12} via ChemShell\cite{she03,met14} to generate the training and test set. The initial training set contained 22 geometries, which corresponds to a total  $2992=22\times (1+15+(15\times 16)/2)$ unique data points, as the system was described with 15 normal modes and the Hessian is symmetric. %The test set initially contained 20 geometries which equals 2720 unique data points.

Using the initial setup several NNs were fitted starting from different initial weights and biases. Then instanton path optimizations were performed on the NNPESs for a large range of temperatures (285K--30K).
Since the instanton path elongates with decreasing temperature and two comparatively high temperature instantons were chosen to define the initial training and test set structures, it was necessary to add further information to the training and test sets to ensure that the section of the NNPESs can describe tunneling correctly over a larger temperature range. Therefore, further structures were chosen along the instantons for temperatures in the medium to lower regime of the temperature range and their energies, gradients and Hessians were added to the training and test set.

Using this improved training and test set a new set of NNs was trained. By iterating the interpolation of NNPESs and adding training structures for regions that are not well described yet by the NNPESs to the training set, improved training and test sets were obtained .The immediate vicinity of the pre-reactive complex is not sufficiently well described by structures chosen along instantons. However, this region must be described very precisely in order to predict reliable unimolecular reaction rate constants. Therefore, energy minimizations starting from the end point of an instanton at the side of the reactant state van-der-Waals minimum were performed. Additional structures along those minimization paths were used for the training and test set. This information improves the description of the pre-reactive complex's vicinity by the NNPESs.  

Our final training set consists of 66 reference structures, which equals to 8976 unique data points. The test set, which is defined by 18 reference structures, contains 2448 data points. 
The final choice of training and test set structures is shown in \figref{fig:instantons} together with instantons at three different temperatures that were calculated on a NNPES that was fitted using the final training and test set. To facilitate the visualization of the structures in \figref{fig:instantons} only their projection on the normal modes 1 and 11 are shown. These modes correspond to the movement of H$_2$, see \figref{fig:modes}. The NN, however, is trained to predict the PES in the complete 15 dimensional space spanned by all normal modes.
The comparatively low number of reference structures that are needed to fit the NNPES describing the reaction at hand is due to the facts that firstly only a local PES is to be fitted and secondly that the information given by the gradients and Hessians allows for a coarser sampling of the PES as they contain information on how the PES will change in the vicinity of the training and test set points.

Thus, in order to find the geometries of the training and test sets, DFT-optimizations of instantons were required to obtain an initial set of reference geometries. All other geometries were obtained by minimizations or from instanton paths calculated on NNPESs. For all these geometries energies, gradients and Hessians were calculated on the UCCSD(T)-level to train and test the NN. This approach is very efficient in CPU time requirements and extendable to larger and more complex reactions.

\subsection{Average NNPES}
In principle every NN fit will lead to a slightly different description of the PES.  This is due to the fact that firstly the initial weights are chosen randomly and secondly the training is done by a local optimization of the weights and biases which can cause the training to converge to different local minima of the residual depending on the starting values of the weights and biases. However, if several NNPESs are available it is a-priori not obvious  which PES is the best approximation of the physically correct PES.
The final values of the cost function $R$ are often very similar for multiple, slightly different NNPESs.

The fact that multiple NNPESs will in general differ slightly in shape also implies that reaction rate constants that are obtained on these surfaces will differ. To obtain a best estimate for the PES and a measure of its local reliability, we averaged several ($N$) NNPESs and used their mean to calculate instantons. Energies, gradients and Hessians were averaged. Their standard error in the energy is used to estimate the reliability of the averaged NNPES for a particular geometry. The standard error $s_{\bar{\text{\tiny{E}}}}$ is defined as
\begin{equation}
	s_{\bar{\text{\tiny{E}}}}(\mathbf{x}) = \sqrt{\frac{\sum_{n=1}^N(E_{\text{\tiny{NN}},n}(\mathbf{x})-\bar{E}_{\text{\tiny{NN}}}(\mathbf{x}))^2}{N(N-1)}}.
\end{equation}
 For a given input structure $\mathbf{x}$, $E_{\text{\tiny{NN}},n}$ denotes the energy of the $n$th individual NN and $\bar{E}_{\text{\tiny{NN}}}$ is the arithmetic average of the $N$ energies predicted for this structure.  
This might also reduce the influence of some local errors that might be contained in a NNPES that otherwise is a good approximation of the PESs because these small local errors can be averaged out if the other NNPESs predict a different course of the PES in these regions. Thus, the averaging serves as a way to regularize the PES and to provide an error estimate. If regions with particularly large $s_{\bar{\text{\tiny{E}}}}$ are reached in the instanton optimizations, additional training points have to be added.

%%%%%%%%%%%%%%%%%%%%%%%%%%%%%%%%%%%%%%%%%%%%%%%%%%%%%%%%%%%%%%%%%%%%%%%%
\section{Results and discussion}
%%%%%%%%%%%%%%%%%%%%%%%%%%%%%%%%%%%%%%%%%%%%%%%%%%%%%%%%%%%%%%%%%%%%%%%%

As an example and to demonstrate the use of NNs including gradient and Hessian information to calculate rate constants, bimolecular reaction rate constants on an average NNPES were calculated for \rxnref{rkn:1}.

\begin{figure}[htbp]
  \centering
  \includegraphics[width = 8cm]{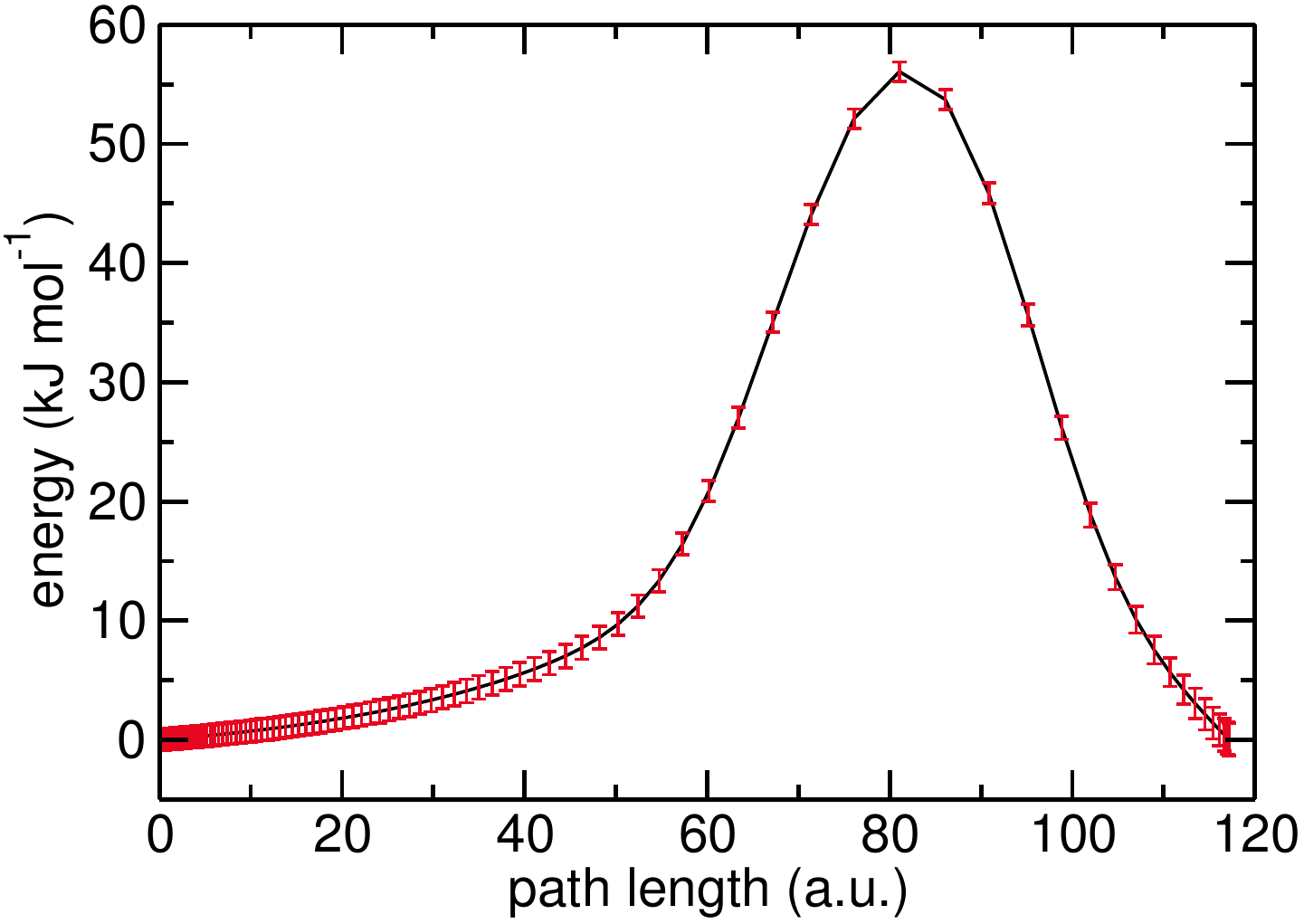}
  \caption{Average potential energy (black) $\pm$ 20 standard errors (red)
    along an instanton at $T=65$~K. The energy is given relative to the energy
    of the first image.}
  \label{fig:energyerror}       
\end{figure}

In \figref{fig:energyerror} the average potential energy along the instanton at 65~K is shown together with its corresponding standard error. The latter was multiplied by 20 to better visualize differences in the standard errors. First it has to be stated that the standard error of the average energy is very small along the whole tunneling path. The maximum standard error is about $0.027$~kJ~mol$^{-1}$, which indicates that the description of the energy along the reaction path is similar for all NNPESs. 

In the vicinity of the transition state structure the standard error is the smallest (0.02~kJ~mol$^{-1}$) and it increases towards the right end of the reaction path, i.e., the product state. This is due to the fact that there are fewer training points in the vicinity of the product state. Thus, there will be a greater variety between individual NNPESs in this region. However, reliable rate calculations are still possible since the error is very small even in these regions of the PES.% very small, the maximum standard error is 0.03 kJ~mol$^{-1}$.

The small error bars of the energy along the whole tunneling path show that the energies that are predicted by the individual NNs at a certain point along the tunneling path are very similar. Therefore, taking an average of these potential surfaces should lead to a regularization of the resulting NNPES. Consequently, performing instanton optimizations on the average surface should yield more reliable rate constants than the individual NNPESs. 

\begin{figure}[t]
  \centering
  \includegraphics[width = 8cm]{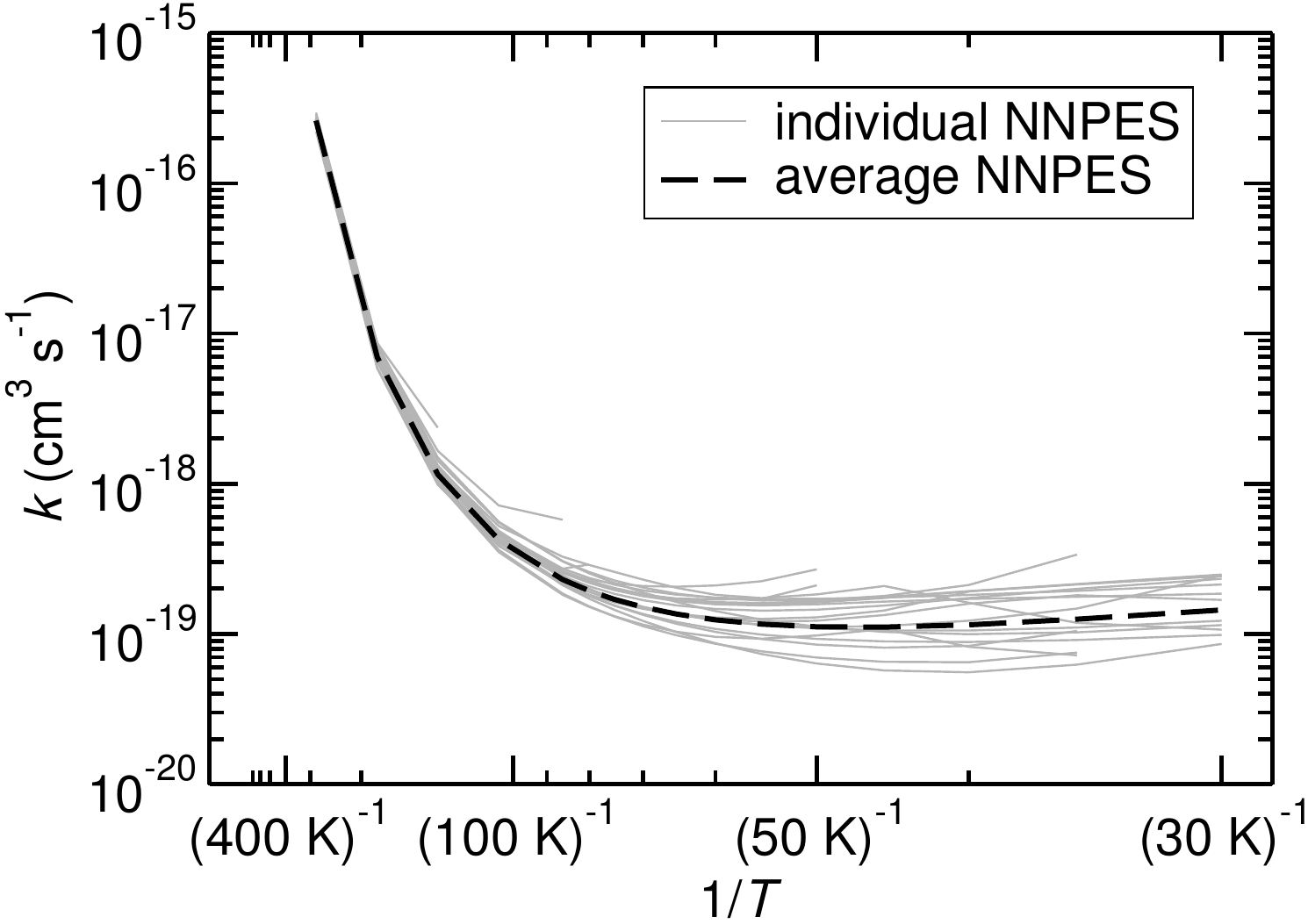}
  \caption{Comparison of bimolecular rate constants for single NNPESs (grey)
    and the rate constant obtained on the average NNPES (dashed, black).}
  \label{fig:rates_single}
\end{figure}

All considerations made so far only covered the potential energy but not its gradient or Hessian, however. In principle it would be possible that that the gradient and especially the Hessian at a specific point might deviate more from the corresponding average than the energy. We decided to test this on the rate constants themselves, rather than using the standard error of gradient or Hessian components.
Therefore, we studied how strongly the rate constant for a given temperature depends on the choice of NNs which contribute to the average NNPES. Rate constants calculated on individual NNPESs, as well as on the average surface, are given in \figref{fig:rates_single}. At low temperature, the individual NNPESs lead to deviations in the rate constants of about a factor of 5 and somewhat uneven temperature-dependences in some cases. The average PES results in a smooth curve.

We compared results for the following three specific selections of NNPESs: 
\begin{description}
\item[Set 1] All NNPESs for which at least one instanton optimization converged  (103 NNPESs).
\item[Set 2] All NNPESs for which the instanton optimization converged for all temperatures tested (63 NNPESs).
\item[Set 3] All NNPESs for which the instanton optimization converged for all temperatures tested and for which a product state geometry could be found (33 NNPESs).
\end{description}

The average over the largest set (1) was chosen to serve as a point of reference because no pre selection of individual NNPESs has to be done.
The second selection of NNPESs ensures that only those hypersurfaces enter the average which describe the shape of the barrier and the vicinity of the pre-reactive complex minimum well as otherwise the instanton optimization would not converge for all temperatures.
The selection of the third set of NNPESs ensures this property as well, but it further ensures that the vicinity of the product state is described by the NNPESs with sufficient accuracy. In sets 1 and 2, there were several NNPESs for which the product channel lead to a deep energy valley rather than a shallow vdW-minimum, which impeded convergence of the product state within the area of configuration space for which the NNPES is reliable. For these three selections of NNPESs we found very similar rate constants for the whole temperature range, see \tabref{tab:rateconst}.

%The instanton optimizations were done for all temperatures discretizing the path to 200 images. \jk{does that mean nimage=200 or nimage=100? The latter would be correct.}

\begin{table}[htbp]
	\caption{Reaction rate constants for different representations of the PES and deviations from the CC reference. All values at $T=65$~K and with 60 images.}
	\centering
	\begin{ruledtabular}
		\begin{tabular}{lclrl}		
			Representation  & Rate constant  & \multicolumn{3}{c}{Deviation from} \\
 			of the PES&[$10^{-19}$ cm$^3$/s]&\multicolumn{3}{c}{the CC reference[$\%$]}\\
			\colrule
			CC reference &2.00&\hspace{0.7cm} &---&\\
			NNPES set 1&2.03&&1.50&\\
			NNPES set 2&1.95&&$-$2.50&\\
			NNPES set 3&2.11&&5.50&\\
		\end{tabular}
	\end{ruledtabular}
	\label{tab:rateconst}
\end{table}

As comparison, we computed one instanton directly with UCCSD(T)-F12/VTZ-F12
on-the-fly. Because of the substantial computational effort involved, we had
to restrict the discretization of the instanton to 60 images. For comparison,
we restricted the NNPES instanton calculation at that temperature to 60 images
as well. The results are given in \tabref{tab:rateconst}. These rate constants
are all very similar. This demonstrates that the NNPES provides very accurate
rate constants compared to on-the-fly calculations. The error from the fit is
restricted to a few percent. The \new{requirement in computational time is
  hugely reduced, though, by about 5 orders of magnitude assuming the the NN
  fit is already available. Overall, 74 Hessians were calculated to fit the
  NNPES (66 for the training set, 18 for the test set). This is comparable to
  the computational requirements of the on-the-fly calculation of one
  instanton rate constant at a single temperature (60 images, 20 optimization
  steps of the instanton), which required 30 Hessian calculations plus
  $20\times 30=600$ gradient calculations. Once the NNPES is fitted, more
  images can be used (here: 200) and instantons at more than one temperature
  and with different isotopologues can be calculated with vanishing additional
  cost.}  The comparison between the NNPES-data also shows that the results
within the different sets of NNPES to be averaged are very similar. Thus, the
largest set 1 will be used in the following. All the errors reported here are
much smaller than the expected intrinsic error of the semiclassical
approximation in instanton theory or the error caused by remaining
inaccuracies of the UCCSD(T) approach.

\begin{figure}[t]
  \centering
  \includegraphics[width = 8cm]{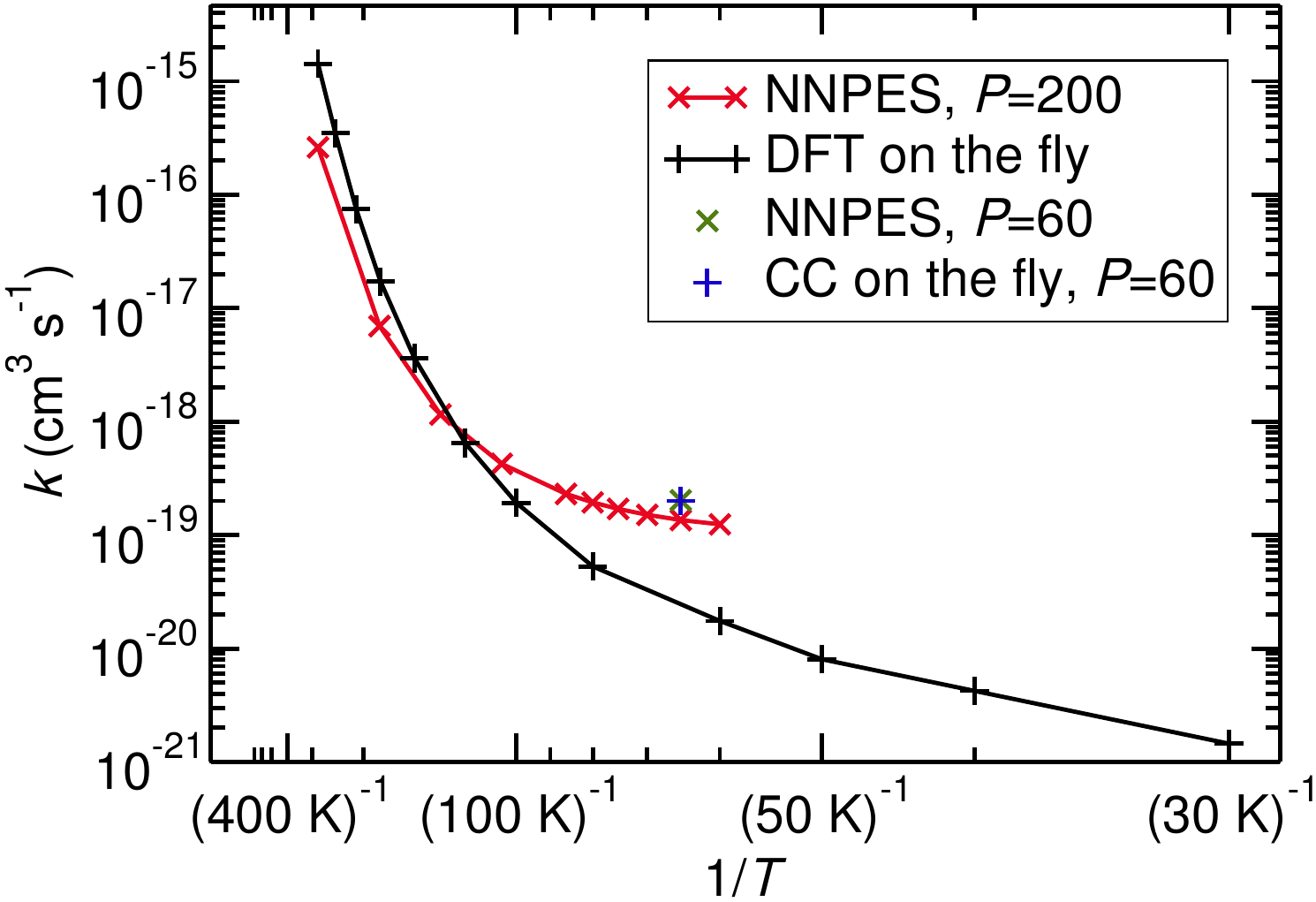}
  \caption{Bimolecular rate constants for an average NNPES fitted on CC level
    (red) and from on-the-fly calculations on DFT level\cite{gou11}
    (black). Reference rate constant obtained from on-the-fly calculations on
    CC level (blue plus sign) and reaction rate constant for the average NNPES
    (green cross).}
  \label{fig:rates}       
\end{figure}

The temperature-dependence of the rate constants that were obtained on the average NNPES with 200 images are shown in \figref{fig:rates}.
The comparison clearly shows qualitative differences to the rate constants on DFT level obtained previously in our group.\cite{gou11} The difference, up to one order of magnitude is caused by the differences in the potential between DFT and coupled cluster. The data using NNPES end at 60~K. Below that temperature, \new{canonical instanton theory becomes inaccurate, because the tunneling energy, the energy of the turning points of the instanton, drops below the energy of the separated species. This can lead to erroneously increasing rate constants at low temperature.}%
Microcanonical instanton theory could be used to extend the temperature range.\cite{mcc17} The pre-reactive vdW-minimum is less deep on the DFT-PES, which allows to calculate rate constants at even lower temperatures.

%##################################################################################################################  
	\FloatBarrier
%%%%%%%%%%%%%%%%%%%%%%%%%%%%%%%%%%%%%%%%%%%%%%%%%%%%%%%%%%%%%%%%%%%%%%%%
\section{Conclusions}
%%%%%%%%%%%%%%%%%%%%%%%%%%%%%%%%%%%%%%%%%%%%%%%%%%%%%%%%%%%%%%%%%%%%%%%%

We have shown that incorporating information on the potential energy, its gradient and Hessian in the NN training yields a NNPES with high accuracy, using relatively few training points. Averaging several NNs trained on the same data further improves the accuracy and yields a measure for the local reliability of the PES fit. The PES is especially suited for rate calculations with instanton theory because it provides smooth and accurate second derivatives. A comparison with on-the-fly calculations of the instanton rate constant demonstrated excellent agreement. This shows that the required CPU time of calculating rate constants can be hugely reduced, by about 5 orders of magnitude, with negligible loss in accuracy. 

\section*{Supporting Information}

Equations for additional derivatives are provided in the supporting material.

\begin{acknowledgments}
This work was financially supported by the European Union's Horizon 2020
research and innovation programme (grant agreement No. 646717, TUNNELCHEM) and
the German Research Foundation (DFG) via the grant SFB 716/C.6. Computational
resources were provided by the state of Baden-W\"urttemberg through bwHPC and
the German Research Foundation (DFG) through grant no INST 40/467-1 FUGG.
\end{acknowledgments}

%%%%%%%%%%%%%%%%%%%%%%%%%%%%%%%%%%%%%%%%%%%%%%%%%%%%%%%%%%%%%%%%%%%%%%%%%
%\section{Misc. info that might turn out to be useful}
%%%%%%%%%%%%%%%%%%%%%%%%%%%%%%%%%%%%%%%%%%%%%%%%%%%%%%%%%%%%%%%%%%%%%%%%%
%
% \begin{itemize}
%	
%	\item the rate constants can be calculated in ca. 4 minutes on my PC in DL-FIND standalone using 1 Processor
%	\item In Chemshell using DL-FIND on Justus with 16 processors on 1 node the on-the-fly CC reaction rate constant needed ca. 5 weeks to be calculated. \
%
%\end{itemize} 

\bibliography{mod_submitted}
 
\vspace{20cm}

\end{document}